\documentclass[a4paper,USenglish,cleveref,autoref]{oasics-v2019}
\usepackage[frozencache,cachedir=.]{minted}

\title{Human-Centric Program Synthesis}

\titlerunning{Human-Centric Program Synthesis}

\author
  {Will Crichton}
  {Stanford University}
  {wcrichto@cs.stanford.edu}
  {}
  {}

\authorrunning{W. Crichton}

\Copyright{John Q. Public and Joan R. Public}


\category{}

\relatedversion{}

\supplement{}

\nolinenumbers 

\hideOASIcs  

\EventEditors{John Q. Open and Joan R. Access}
\EventNoEds{2}
\EventLongTitle{42nd Conference on Very Important Topics (CVIT 2016)}
\EventShortTitle{CVIT 2016}
\EventAcronym{CVIT}
\EventYear{2016}
\EventDate{December 24--27, 2016}
\EventLocation{Little Whinging, United Kingdom}
\EventLogo{}
\SeriesVolume{42}
\ArticleNo{23}

\begin{document}

\maketitle

\begin{abstract}
Program synthesis techniques offer significant new capabilities in searching for programs that satisfy high-level specifications. While synthesis has been thoroughly explored for input/output pair specifications (programming-by-example), this paper asks: what does program synthesis look like beyond examples? What actual issues in day-to-day development would stand to benefit the most from synthesis? How can a human-centric perspective inform the exploration of alternative specification languages for synthesis? I sketch a human-centric vision for program synthesis where programmers explore and learn languages and APIs aided by a synthesis tool.
\end{abstract}

\section{A Story of Our Time}

Consider the story of Dana the Data Scientist. At Sonmanto, her agritech business, Dana wants to analyze the weekly seed production. Opening a Jupyter notebook, she creates a new code cell, imports a few libraries, and sends off a SQL query to build a Pandas dataframe. Knowing the Pandas API from her data science course and past experience, she computes the week's average seed production using standard dataframe methods.
\begin{minted}{python}
query = sql("SELECT time, production FROM seed_production ORDER BY time")
df = pd.read_sql_query(query)
df.where(df.time >= dt.now() - dt.timedelta(days=7)).production.mean()
\end{minted}
Concerned that the week's production seems low, Dana wants to see a 7-day rolling  average of the last year's production to put this week into context. She has never computed a weekly rolling average before, so she Googles ``pandas rolling average''. Excellent, Pandas has a \verb|Dataframe.rolling| method, but\ldots\ it doesn't do quite what she wants. All the examples use windows that contains a fixed number of elements, but she wants windows of a fixed duration potentially containing different numbers of production samples. 

Dana continues searching increasingly elaborate queries like ``pandas rolling average date dynamic window'', and eventually finds some StackOverflow answers that look almost right. However, all of their solutions either use abstract notation like ``foobar'' or were made for other domains like stock trading. Dana finds it difficult to see the relationship between finance problems and seed production. After twenty minutes of searching, she gives up with a resigned sigh and decides to just implement it in plain Python.
\begin{minted}{python}
for day_start in pd.date_range(df.time.min(), df.time.max()):
    day_end = day_start + datetime.timedelta(days=7)
    window = [row.production 
              for day in pd.date_range(day_start, day_end) 
              for _, row in df.iterrows() if row.time.date() == day.date()]
    weekly_prod.append(pd.Series(window).mean())
\end{minted}
Dana knows the code isn't beautiful, but it gets the job done. Glancing back at the StackOverflow questions, she starts to see the connections after going through her own implementation. But it's close to lunch, and she spent too long on this already. Simplifying the code is a task for another day, and she moves on.

\section{A Story of Another Time}

\ldots Concerned that the week's production seems low, Dana wants to see a 7-day rolling  average of the last year's production to put this week into context. She has never computed a weekly rolling average before, so she Googles ``pandas rolling average''. Excellent, Pandas has a \verb|Dataframe.rolling| method, but\ldots\ it doesn't do quite what she wants.

Rather than continuing to search fruitlessly, she writes down her plain Python solution. Dana highlights the code cell and clicks ``Synthesize'' in her Jupyter toolbar, opening a dialog box on the side. She writes \verb|Dataframe.rolling| and \verb|Dataframe.mean| into the box, knowing those are likely going to be important parts of a Pandas-specific solution if it exists. Guided by her suggestions, the synthesis engine finishes in under a minute, producing a \verb|rolling| solution contextualized to her dataframe.
\begin{minted}{python}
df.sort_values('time').set_index('time').rolling('7d').mean()
\end{minted}
Ahh, the rolling function has a special syntax for time windows. But, Dana wonders, what does each part do? Hovering over each part of the program, the synthesis tool uses its counterexamples to shows what would happen if a given method call was omitted or changed. Removing \verb|sort_values| or \verb|set_index| cause the program to raise an error. Changing the window to \verb|rolling(7)| produces an incorrect output.

Plotting the values in Matplotlib, Dana marvels at the simplicity of the solution. She starts to wonder: are there other places in Sonmanto's code base where they could use this pattern? Glancing over at the clock, there's still an hour to lunch, great! Highlighting her old code cell once more, she clicks ``Find Similar'' to search her notebooks and text files for snippets that look structurally similar to the one she just wrote. 

After the search engine returns five plausibly similar programs, Dana runs the synthesis engine in parallel on each one. Noticing that most of the snippets were written by Danny the Data Engineer, she motions Danny over and teaches him about the feature she just learned.

\section{The Past and Present of Program Synthesis}

The stories above highlight a key fact about modern-day programming: programmers routinely deal with dozens of representations of code and data. In the data science domain, Jupyter notebooks swap between explanation and code. Data flows from SQL databases to Python lists to Pandas dataframes. Operations mix and match bespoke APIs with general-purpose programming constructs. Programmers are continually learning new representations as languages, libraries, and tools emerge and change. 

Dana's struggles with Pandas show a prototypical case of acquiring a new representation. Not knowing how to compute a specific kind of rolling average, she uses a combination of documentation, code examples, and prior knowledge to understand whether the Pandas API can solve her problem. Having general programming skills, she can arrive at a standard Python solution, but not the more concise API-specific solution. As the second story demonstrates, I believe that program synthesis techniques hold promise in helping programmers overcome these kinds of representational transfer problems (or refactoring, migration, etc.). Yet, to date, such a story is still a fantasy.

To understand why, we will briefly examine the history of program synthesis. Synthesis has been predominantly applied in the context of programming-by-example (PBE). In PBE, a user provides examples (input/output pairs) of a pure function, and the synthesizer attempts to find a ``good'' (e.g. small) function that satisfies those examples. Often, the user is an end-user manipulating spreadsheets or text documents, and the generated program is an invisible macro. Through hard-earned experience with dozens of PBE systems, researchers have both articulated design principles of PBE\,\cite{lau2009programming} and ultimately produced the flagship commercial synthesis engine, Excel FlashFill\,\cite{polozov2015flashmeta}. This effort succeeded in part by a human-centric push to understand both the applications where PBE was most valuable, and the essential usability constraints for real-world usage.

The central question of this paper, then: what does human-centric synthesis look like beyond PBE? Specifically, what applications open up when a user has the programming skills to express specifications at a level beyond examples? Traditionally, these kinds of tasks have been viewed as refactoring or migration, where the existing codebase specifies the desired behavior for the transformed one. Historically, refactoring tools could only perform simple syntactic changes like renaming types or methods. However, recent synthesis tools have shown striking progress in translating between complex high-level representations of code. For example, tools can move between languages like Java $\rightarrow$ Spark\,\cite{ahmad2018automatically}, and Fortran $\rightarrow$ Halide\,\cite{kamil2016verified}. Tools can refactor APIs, like parallelizing Java streams\,\cite{khatchadourian2019safe}, adding default methods in Java\,\cite{khatchadourian2017automated}, updating SQL queries after schema changes\,\cite{wang2019synthesizing}.

These approaches have significantly advanced the state-of-the-art in program synthesis techniques. But given the lack of meaningful commercial adoption, it is unclear whether they're trying to solve the right problem. Certainly this fact arises in part from the general difficulty of tech transfer. But this would not be the first time the PL and compilers communities have been led astray by the allure of automating high-level tasks for programmers. 

For example, modern compilers do register allocation by solving a complex graph coloring problem with zero user input, and no one takes issue with that. History and experience suggest that deciding which temporary is assigned to which register or stack slot is not a meaningful decision for a programmer. By contrast,  decades of research have been invested into automatic parallelization of general-purpose code, like arbitrary C for-loops. However, identifying and expressing parallelism in an application seems more fundamental/important for the programmer to decide than register allocation. Automatic techniques have been ultimately eclipsed by DSLs with understandable, user-programmable models of parallelism like Halide\,\cite{ragan2013halide} and Spark\,\cite{zaharia2010spark}. After all, autovectorization is not a programming model\,\cite{pharr}. So how can program synthesis avoid a similar fate?

\section{A Vision for Human-Centric Program Synthesis}

The moral of these stories is that understanding the context, challenges, and capabilities of the working programmer is essential for improving the programming experience. Applying such a human-centric lens in designing and evaluating synthesis tools could accelerate the progress of synthesis research and promote the real-world adoption of these techniques. While the goal of this paper is primarily to spark discussion---what do you think human-centric synthesis entails?---I want to set the stage by articulating my principles for improving synthesis tools.

\begin{enumerate}
    \item \textbf{Synthesis tools should use a user's most productive specification language.}  
    
    Input/output pairs have been a popular specification language for synthesis, since "PBD is a natural match for artificial intelligence... by observing the actions taken by the user (training examples), the system can create a program (learned model) that is able to automate the same task in the future."\,\cite{lau2009programming} Moreover, for end-users without training in formal languages, I/O pairs are the highest level of abstraction at which they can formally specify behavior. However, programmers can use a diverse array of representations for specifications. These range from testing (e.g.\ unit testing, randomized test generation\,\cite{claessen2011quickcheck}) to declarative languages (PlusCal, UML) to programming languages (sketches\,\cite{solar2006combinatorial}).
    
    Synthesis tools should use a specification language based on the difficulty of writing abstract rules versus writing individual examples in a given domain. Dana's window query may be easier for her to specify in Python, while a data structure manipulation like rotating a tensor may be easier to specify by examples. Human-centric evaluations of synthesis should seek to empirically characterize this trade-off. \\
    
    \item \textbf{The synthesized program can not be a black box.} 
    
    Synthesis tools have historically been used like compilers: input the specification, and don't look at the output program, just run it. Again, while this approach works for end-users who may lack the technical knowledge to understand the synthesized program, such an interaction mode is rarely desirable for a programmer. Programs are written, re-read, tweaked, maintained, handed off to other collaborators, and so on. Professional programmers spend only 5\% of their time writing code\,\cite{minelli2015know,xia2017measuring}. 
    
    Subsequently, Programmers must be able to comprehend and maintain synthesized programs. A synthesis tool should generate readable code and be able to explain its decisions, like Dana's imagined UI. Readability metrics can be informed by existing principles of programming notation design, like the cognitive dimensions framework\,\cite{green1989cognitive}.
\end{enumerate}

\noindent These principles have helped me envision application spaces beyond the traditional purview of synthesis, like those characterized in the stories above. For example:

\begin{enumerate}
    \item \textbf{Helping programmers learn new languages and APIs.} 
    Programmers, whether hobbyists or full-time developers, encounter learning opportunities every time they code. Dana's was intentional: she realized she didn't know a feature and searched for it. But many more opportunities are passed by due to lack of awareness of a language or API feature. Anecdotally, I know Rust users that say the linter (Clippy) has helped them learn APIs through simple syntactic patterns. A synthesis tool as an extremely powerful linter could identify when someone likely doesn't know a concept (``there are 10 places in your code base that could be simplified with a for-loop''), highlight relevant code, and even suggest the translation if possible. By explaining its code-generating decisions, a synthesis tool can move beyond code that just works, to code that teaches how it works. \\
    
    \item \textbf{Evaluating the impact of an API/language change.} When maintainers of libraries and languages debate new features, questions arise like: how many people would use this change? Would their code be meaningfully improved with this feature? For example, the Python community recently accepted the contentious PEP 572 ``walrus operator.'' Guido was ultimately convinced by maintainers who combed through their own codebase, demonstrating dozens of places where the proposed feature could be applied\,\cite{lwn}. A synthesis tool could help maintainers automate such exploratory tasks and more freely experiment with proposed designs.
\end{enumerate}

\noindent Enabling these applications raises a number of exciting research questions in the design, implementation, and evaluation of synthesis tools. If high-level specifications replace I/O pairs, does this reduce the program search space, or is it just a means of generating examples (like QuickCheck)? Can API or language designers make their systems more amenable to synthesis? I hope that perspectives from the PL/HCI community can contribute greatly to these endeavors.

\section{Acknowledgments}

I would like to thank my advisor Pat Hanrahan for his everlasting support despite my constantly evolving research direction. And a big thanks to Brian Hempel, Georgia Gabriela Sampaio, and my anonymous reviewers for constructive comments that substantially improved the quality of this paper.

\bibliography{oasics-v2019-sample-article}

\end{document}